\documentclass[12pt]{article}

\usepackage{moreverb}

\usepackage[colorlinks,bookmarksopen,bookmarksnumbered,citecolor=red,urlcolor=red]{hyperref}

\newcommand\BibTeX{{\rmfamily B\kern-.05em \textsc{i\kern-.025em b}\kern-.08em
T\kern-.1667em\lower.7ex\hbox{E}\kern-.125emX}}

\usepackage{graphicx}
\usepackage{natbib}
\usepackage{lscape,rotating}
\usepackage{float,epsfig}
\usepackage{subfigure}
\usepackage{setspace}
 \usepackage{amsmath,bm} 
\usepackage{amsfonts}
\usepackage{amssymb}

\newtheorem{example}{Example}

\newcommand{\zv}{\bm{z}}

\newcommand{\alphav}{\bm{\alpha}}

\newcommand{\gammav}{\bm{\gamma}}
\newcommand{\lambdav}{\bm{\lambda}}

\newcommand{\zerov}{\bm{0}}

\begin{document}

\def\spacingset#1{\renewcommand{\baselinestretch}%
{#1}\small\normalsize} \spacingset{1}

\spacingset{1.45} 


\title {\bf An improved sample size calculation method for score tests in generalized linear models }
\author{Yongqiang Tang \footnote{email: yongqiang\_tang@yahoo.com} \\
  Tesaro, 1000 Winter St, Waltham, MA 02451 \\
  Liang Zhu \\
    The University of Texas Health Science Center at Houston, Houston, TX 77030\\
    Jiezhun Gu\\
  Duke Clinical Research Institute, Durham, NC 27705}
\maketitle

\bigskip

\begin{abstract}
\cite{self:1988} developed a sample size determination procedure for score tests in generalized linear models under contiguous alternatives. 
Its performance may deteriorate when the effect size is large. We propose a modification of the Self-Mauritsen method by taking into account of the variance of the score statistic
 under both the null and alternative hypotheses, and extend the method to noninferiority trials. 
The modified approach is employed to calculate the sample size for the logistic regression and  negative binomial  regression in 
superiority and noninferiority trials.  We further explain why the formulae recently derived by Zhu and Lakkis tend to underestimate the required sample size for the  negative binomial  regression. 
Numerical examples are used to demonstrate the accuracy of the proposed method. 
\end{abstract}
\noindent%
\vfill

\section{Introduction}
Generalized linear models  (GLM) have been commonly used in the analysis of biomedical data \citep{nelder:1972,McCullagh:1989}. Statistical inference in GLMs is often based on 
the Wald test and the likelihood ratio (LR) test. However, the Wald and LR tests can be liberal in small  and moderate samples.  
In the comparison of two binary proportions, the Wald and LR methods can be anti-conservative under some parameter configurations even when the sample size reaches $200$   \citep{laud:2014}
because the logistic regression overestimates the odds ratio in these studies  \citep{nemes:2009}. 
Similar phenomenon is observed in the analysis of over-dispersed count data using the negative binomial (NB) regression \citep{aban:2008}. 
The score test has been recommended to control the type I error rate when the sample size is relatively small. 
In fact, many widely used methods such as Pearson's chi-squared test, Cochran-Mantel-Haenszel test and Wilcoxon rank sum test are score tests from GLMs.

One concern about the score test is its lower power when compared to  the Wald test. In fact,
the score test can sometimes be more powerful than the Wald test. \cite{xing:2012} observed that the Wald test from the  logistic regression may
often miss  rare disease-causal variants that can be identified by other asymptotic tests in large case-control association studies. 
Table \ref{scorehigh} presents  two scenarios for comparing two binomial proportions on the risk difference metric, in which the score test  has higher power than the Wald test.
The first scenario tests for superiority when the sample sizes are unbalanced in the two groups.
In scenario 2, a noninferiority (NI) test is considered under balanced sample sizes. 
The results also evidence that it may sometimes be inappropriate to use the power calculation procedure developed for the Wald test to estimate the power of the score test, and  vice versa.
Technical details on the score test and the exact power calculation can be found in  \cite{farrington:1990} and \cite{tang:2019b}.

\begin{table}[h]
\begin{center}
\small
\begin{tabular}{c@{\extracolsep{5pt}}c@{\extracolsep{5pt}}c@{\extracolsep{5pt}}c@{\extracolsep{5pt}}c@{\extracolsep{5pt}}c@{\extracolsep{5pt}}c@{\extracolsep{5pt}}c}\\\hline 
\multicolumn{2}{c}{group size} & \multicolumn{2}{c}{true proportion}  && \multicolumn{3}{c}{exact power ($\%$)}\\ \cline{1-2}\cline{3-4} \cline{6-8}
$n_1$ & $n_0$ & $p_1$ & $p_0$ & Hypothesis & score$^{(a)}$ & Wald$^{(b)}$ & Wald2$^{(b)}$ \\\hline
60 & 30 & 0.1 & 0.3 & $H_0: p_1\geq p_0$ {\it vs } $H_1: p_1<p_0$ & 67.33 & 60.81 & 65.28\\
80 & 80 & 0.35 & 0.4 &  $H_0: p_1\geq p_0+0.15 $ {\it vs } $H_1: p_1<p_0+0.15$ & 75.37 & 74.05& - \\
\hline
\end{tabular}\caption{Two scenarios with higher power in the score test than in the Wald test\newline
$^{(a)}$ Score test defined in equation (3) of  \cite{farrington:1990} \newline
$^{(b)}$ Wald ($Z= \frac{\hat{p}_1-\hat{p}_0- M_0}{\sqrt{\hat{p}_1(1-\hat{p}_1)/n_1+ \hat{p}_0(1-\hat{p}_0)/n_0}}$) and Wald2 ($Z= \frac{\log\{[\hat{p}_1/(1-\hat{p}_1)]/[\hat{p}_0/(1-\hat{p}_0)]\}}{\sqrt{[n_1\hat{p}_1(1-\hat{p}_1)]^{-1}+ [n_0\hat{p}_0(1-\hat{p}_0)]^{-1}}}$)
are the Wald tests from the binomial regression respectively with identity and logit link functions
}\label{scorehigh} 
\end{center}
\end{table}
 
 \cite{self:1988} developed a power and sample size calculation procedure for the score test from GLMs under
sequences of contiguous alternatives \citep{cox:1974}. This method generally works well for alternatives close to the null hypothesis. Its accuracy may degrade when the group sample sizes are unbalanced or when
the effect size is large \citep{self:1992}. 
Self and Mauritsen's approach approximates  the variance of the score statistic under the null hypothesis  by the variance under the alternative hypothesis.
This assumption is asymptotically correct under contiguous alternatives, but unlikely to hold at alternatives that are not close to the null hypothesis \citep{self:1988}. 

We propose a  modification of Self and Mauritsen's procedure by taking into account of the variance of the score statistic
 under both the null and alternative hypotheses. It can greatly improve the performance of the method.
For example, \cite{tang:2011}  obtained the sample size formula for Wilcoxon rank sum test for ordinal outcomes  on basis of the asymptotic variance of the U statistic under both hypotheses, which shows improvements over the formulae derived under contiguous alternatives  \citep{whitehead:1993, zhao:2008}.
Similar ideas were employed by  \cite{farrington:1990}  in the comparison of binary proportions in NI trials.
In these simple cases, the score test and its asymptotic distribution can be obtained analytically. In this paper, we consider  more complex situations where the model contains some nuisance parameters. 
The score test has been commonly used in the superiority trials.
It is less well known  how to use the score method to analyze the NI trials. 
In   Section \ref{scoreglm}, we explain how to conduct the  NI tests in GLM based on the score method, and  
introduce the modified  sample size procedure for both superiority and NI trials via the exemplary dataset approach.

The proposed method is employed to estimate the sample size  for the  score test from the NB regression  in Section \ref{nbscore}, and
for the score test from the logistic regression with categorical covariates in Section \ref{logissec}.  The performance of the proposed method is assessed by numerical examples and compared with some existing procedures. 

\section{Score tests in GLM}\label{scoreglm}
\subsection{Score test and score confidence interval}
In GLMs, the scalar response variables $y_1,\ldots,y_n$ are assumed to have probability density functions of the form \citep{nelder:1972,McCullagh:1989} 
\begin{equation}\label{modelj}
 f(y_{i}|x_i,\zv_i,\beta,\alphav,\phi) = \exp\left[\frac{y_{i} \theta_i-b(\theta_i)}{a(\phi)} + c(y_{i},\phi)\right],
\end{equation}
where  $\theta_i$ is the  canonical parameter and $\phi$ is the dispersion parameter. The mean of $y_i$ is  $\mu_{i}=\frac{\partial b(\theta_i)}{\partial\theta_i}$, and its variance is $V_{i}=\frac{\partial \mu_{i}}{\partial\theta_i }a(\phi)$. 
We assume that  the covariates are related to the mean  $\mu_{i}$ via a link function 
$\eta_{i}= \beta x_i+ \alphav'\zv_i=g(\mu_{i})$, where $x_i$ is a scalar covariate, the vector $\zv_i$ contains other covariates including the intercept,
and $(\beta, \alphav)$ are the regression coefficients. In the analysis of clinical trials, $x_i=0$ or $1$ is the treatment status.

Suppose we are interested in testing the hypothesis 
$$H_0: \beta=\beta_0 \text{ {\it vs} } H_1: \beta\neq\beta_0.$$ 
Let
$ \begin{bmatrix}  S_{\beta_n}(\beta,\lambdav)  \\ S_{\lambdav_n}(\beta,\lambdav)  
\end{bmatrix}=\begin{bmatrix} \sum_{i=1}^n \frac{\partial\log[f(y_{i}|x_i,\zv_i,\beta,\lambdav)]}{\partial \beta }  \\ \sum_{i=1}^n \frac{\partial\log[f(y_{i}|x_i,\zv_i,\beta,\lambdav)]}{\partial\lambdav }  
\end{bmatrix}$ 
 be the score function, and  \\
$I_{n}(\beta,\lambdav) =\begin{bmatrix} I_{\beta\beta_n}(\beta,\lambdav) & I_{\beta\lambdav_n}'(\beta,\lambdav) \\
              I_{\beta\lambdav_n}(\beta,\lambdav) & I_{\lambdav\lambdav_n}(\beta,\lambdav) \\\end{bmatrix} $ the expected information matrix,
where the subscript $_n$ refers to the sample size, and $\lambdav =(\alphav',\phi)$ is the vector of nuisance parameters.
Let $\hat\lambdav=(\hat\alphav,\hat\phi)$ be the maximum likelihood estimate (MLE) under the restriction of $\beta=\beta_0$. That is, $n^{-1}S_{\lambdav_n}(\beta_0,\hat\lambdav)=\zerov$.

The score statistic for testing $H_0: \beta=\beta_0$ can be written as \citep{cox:1974}
\begin{equation}\label{scoretest}
 Z(\beta_0) = \frac{S_{\beta_n}(\beta_0,\hat\lambdav) }{ \sqrt{V_n}}
= \frac{ S_{\beta_n}(\beta_0,\hat\lambdav) }{ \sqrt{I_{\beta\beta_n}(\beta_0,\hat\lambdav) -I_{\beta\lambdav_n}'(\beta_0,\hat\lambdav) I_{\lambdav\lambdav_n}^{-1}(\beta_0,\hat\lambdav) I_{\beta\lambdav_n}(\beta_0,\hat\lambdav)}}.
\end{equation} 
 The null hypothesis is rejected if $|Z(\beta_0)|\geq z_{1-\alpha/2}$, where $z_p$ is the $p$th percentile of $N(0,1)$.

As will be illustrated in Section \ref{nbscore}, the score test can be used to test the hypothesis in superiority and NI trials by setting $\beta_0$ as the superiority and NI margin. The confidence interval (CI) is often reported to quantify the uncertainty in the estimated effect. 
The $(1-\alpha)100\%$ score CI for $\beta$  can be obtained by inverting the score test
  $$\{\beta: |Z(\beta)|\leq z_{1-\alpha/2}\}.$$
Statistical decision can be made equivalently based on the score CI.  The null hypothesis is rejected if the score CI does not contain the null hypothesis value.

\subsection{Asymptotic distribution of the score statistic}\label{asyscore}
The score test and its asymptotic distribution usually have explicit analytic expressions in the simple two-group comparison if the model does not contain an unknown dispersion parameter. 
Please refer to  \cite{farrington:1990} and \cite{tang:2019b} for examples. We consider  more general cases where  the vector of nuisance parameters $\lambdav$ contains other parameters in addition to an intercept term.

In general, $\hat\lambdav$ is not a consistent estimate of $\lambdav$ under the restriction of $\beta=\beta_0$. It will converge to the limiting value $\lambdav^*$ defined as the solution to the following equation \citep{self:1988}
\begin{equation}\label{scoreinfty}
  \lim_{n\rightarrow\infty} \text{E}[n^{-1}S_{\lambdav_n}(\beta_0,\hat\lambdav)] =\zerov.
\end{equation}

We estimate $\lambdav^*$ by adapting the method of  \cite{lyles:2007}. We firstly construct an exemplary dataset 
consisting of records for every possible combination of the covariates and outcomes. Each record has a weight that  represents
the frequency of the covariate and outcome in the population. A weighted regression is fitted to the  exemplary dataset using the standard statistical software.
We assume that all covariates  are categorical. A continuous covariate can be discretized using a large number of categories. 
Suppose there are a finite number of distinct covariate configurations $\{(\zv_k, x_k); k = 1, ..., m\}$, and the proportion of each configuration is $\pi_k$ in the population.
Suppose the response variable takes $J$ possible values $(y_1, \ldots, y_J)$.  We can estimate $\lambdav^*$ by fitting the null model to   
the following dataset with $mJ$ observations, where $w_{kj}= \pi_k \Pr(Y=y_j|x_k,\zv_k,\beta,\lambdav)$ is the weight attached to each observation, and  the total weight in all observations is $\sum_{i=1}^m \sum_{j=1}^J w_{ij}=1$. 
\begin{table}[h]
\begin{center}
\begin{tabular}{cccrccccccrccccccc}
$x_1$ & $\zv_1$ & $y_1$ & $w_{11}$\\
$\ldots$\\
$x_1$ & $\zv_1$ & $y_J$ & $w_{1J}$\\
$\ldots$\\
$x_m$ & $\zv_m$ & $y_1$ & $w_{m1}$\\
$\ldots$\\
$x_m$ & $\zv_m$ & $y_J$ & $w_{mJ}$\\
\end{tabular}\end{center}
\end{table}

\newpage

 \cite{lyles:2007} approach is slightly different. It requires a much larger dataset, and can only estimate the power at a given sample size.
The total weight in  \cite{lyles:2007} approach is equal to the total sample size $N$.  
Let's give a simple example of comparing two binary proportions with $\Pr(y=1|x=1)=0.8$ and $\Pr(y=1|x=0)=0.4$. Suppose $N=100$, and $75\%$ patients are assigned to the experimental arm $(x=1)$.
Then $(m,J)=(2,2)$. Our approach includes $4$ observations:
 $(x=1, y=1,w=0.6)$,  $(x=1, y=0, w=0.15)$,  $(x=0, y=1,w=0.1)$ and  $(x=0, y=0, w=0.15)$.   
In \cite{lyles:2007} approach, the dataset consists of $200$ pseudo-observations with $75$ copies of $(x=1, y=1, w= 0.8)$ and $(x=1, y=0, w= 0.2)$ for  subjects in the experimental arm,
and $25$ copies of $(x=0, y=1, w= 0.4)$ and $(x=0, y=0, w= 0.6)$ for placebo subjects. 
If observations with the same  $(x,y)$ are combined by adding up their weights, the dataset  in  \cite{lyles:2007} approach  becomes a dataset with four observations 
 $(x=1, y=1,w=60)$,  $(x=1, y=0, w=15)$,  $(x=0, y=1,w=10)$ and  $(x=0, y=0, w=15)$.
The ratio of the weights for observations with the same $(x,y)$ is $N:1$  between \cite{lyles:2007} approach and our approach.
 
 Our method is more convenient and potentially more accurate than \cite{lyles:2007} approach. 
In \cite{lyles:2007} method, one needs to guess the sample size, construct the exemplary dataset, fit the null model and estimate 
the power at the given sample size. The whole process needs to be repeated if the sample size changes.  In theory,  $\lambdav^*$ remains unchanged, and the noncentrality parameter of the 
Wald, Score or LR test  or its square change proportionally if we  increase or decrease the total sample size. Therefore, the power and sample size calculation can be implemented 
by first fitting the model at a fixed sample size, and then using analytic methods to adjust the  noncentrality parameter
and  solve the power or sample size equations accordingly. 
We fit the null model using SAS Proc Genmod with the FREQ option to incorporate the weight. 
One shall not use the Weight option in the Genmod procedure since it is used to adjust for the dispersion parameter.
Because the Genmod procedure truncates the weight to an integer, we {\color{red} multiply} all the weights by a large value (say $10^9$) to minimize the effect of truncation.  
\cite{lyles:2007} fits the model at the sample size for the trial, which is typically small (i.e.  below $1,000$). The weight after truncation in \cite{lyles:2007} method may no longer represent 
the frequency of the covariate and outcome in the population, and  the estimation of $\lambdav^*$ can be inaccurate.

 In \cite{self:1988}, an exemplary dataset contains $m$ records   for all possible combinations of the covariates,
where the weight is the frequency of the covariates, and the response outcome is the expected value of response at the covariate configuration.
The data structure may not be acceptable by some statistical software packages, and does not allow the estimation of  the dispersion parameter $\phi^*$.

In GLMs, the inference is made by assuming the covariates are known and fixed, but the covariates are typically unobserved at the design stage of a clinical trial.
For example, although gender  is fixed for each patient, it will be treated as unknown at the design stage  since we do not know which patients will be enrolled. 
We firstly derive the mean and variance of the score statistic given the covariates, which are then averaged over all possible combinations of the covariates.

Let $S^{(k)}(\beta,\lambdav)= \begin{bmatrix}  S_{\beta}^{(k)}(\beta,\lambdav)  \\ S_{\lambdav}^{(k)}(\beta,\lambdav)  
\end{bmatrix}=\begin{bmatrix}  \frac{\partial\log[f(y_i| x_k, \zv_k, \beta,\lambdav)]}{\partial \beta}\\
 \frac{\partial\log[f(y_i| x_k, \zv_k, \beta,\lambdav)]}{\partial\lambdav }  
\end{bmatrix}$  denote the contribution to the score function from a subject with covariate $(x_k,\zv_k)$.
Let $E^{(k)}$ and $V^{(k)}$ represent, respectively, the mean and variance of  $S^{(k)}(\beta_0,\lambdav^*)$ under the true model \eqref{modelj}.
Let $E=(E_\beta,E_\lambda)'=\sum_{k=1}^m \pi_k E^{(k)}$ and $V=\sum_{k=1}^m \pi_k V^{(k)}$.  Note that $E_\lambda=\zerov$ by equation \eqref{scoreinfty}.
The asymptotic distribution of   $(S_{\beta_n}(\beta,\lambdav^*) , S_{\lambdav_n}(\beta,\lambdav^*) )'$ is given by 
\begin{equation}\label{var0score0} 
   n^{1/2}\begin{bmatrix} n^{-1} S_{\beta_n}(\beta,\lambdav^*) -E_\beta, \\ n^{-1} S_{\lambdav_n}(\beta,\lambdav^*)-E_\lambda\end{bmatrix} \sim N(0, V).
\end{equation}

Let $\mathcal{J}^{(k)}(\beta,\lambdav)   =-\begin{bmatrix} \frac{\partial^2\log[f(y_i| x_k, \zv_k, \beta,\lambdav)]}{\partial \beta^2 }  &    \frac{\partial^2\log[f(y_i| x_k, \zv_k, \beta,\lambdav)]}{\partial \beta\partial\lambdav' }    \\
\frac{\partial^2\log[f(y_i| x_k, \zv_k, \beta,\lambdav)]}{\partial \beta\partial\lambdav }  &    \frac{\partial^2\log[f(y_i| x_k, \zv_k, \beta,\lambdav)]}{\partial\lambdav\partial\lambdav' }  \\
\end{bmatrix}$ denote  the contribution to the  observed information matrix from a subject with covariate $(x_k,\zv_k)$.
Let $\tilde{I}^{(k)}(\beta_0,\lambdav^*)$  be the expectation of $\mathcal{J}^{(k)}(\beta_0,\lambdav^*)$  under the true model
 \eqref{modelj}, and   $\tilde{I}(\beta_0,\lambdav^*)=\begin{bmatrix} \tilde{I}_{\beta\beta}(\beta_0,\lambdav^*) & \tilde{I}_{\beta\lambdav}'(\beta_0,\lambdav^*) \\
              \tilde{I}_{\beta\lambdav}(\beta_0,\lambdav^*) & \tilde{I}_{\lambdav\lambdav}(\beta_0,\lambdav^*) \\\end{bmatrix}
=\sum_{k=1}^m \pi_k \tilde{I}^{(k)}(\beta_0,\lambdav^*)$.
By the  Taylor series expansion, we get
\begin{equation}\label{taylor4}
 n^{-1/2}S_{\beta_n}(\beta_0,\hat\lambdav)  \approx n^{-\frac{1}{2}} S_{\beta_n}(\beta_0, \lambdav^*) - \tilde{I}_{\beta\gammav}'(\beta_0,\gammav^*)  \,\tilde{I}_{\gammav\gammav}^{-1}(\beta_0,\gammav^*)  n^{-\frac{1}{2}} S_{\gammav_n}(\beta_0, \gammav^*).
\end{equation}

Combining equations \eqref{var0score0} and \eqref{taylor4} yields the asymptotic distribution of the score statistic 
\begin{equation}\label{distscore}
n^{\frac{1}{2}} [n^{-1}S_{\beta_n}(\beta_0,\hat\lambdav) - E_\beta] \sim N(0, \sigma_1^2),
\end{equation}
where $ \sigma_1^2 = A V A'$, $A=[1,  - \tilde{I}_{\beta\gammav}'(\beta_0,\gammav^*)  \,\tilde{I}_{\gammav\gammav}^{-1}(\beta_0,\gammav^*)]$.
In the special case considered by  \cite{self:1988}, $\tilde{I}(\beta_0,\lambdav^*)$ is identical to the Fisher information matrix
$I(\beta_0,\lambdav^*)=\begin{bmatrix} I_{\beta\beta}(\beta_0,\lambdav^*) & I_{\beta\lambdav}'(\beta_0,\lambdav^*) \\
              I_{\beta\lambdav}(\beta_0,\lambdav^*) & I_{\lambdav\lambdav}(\beta_0,\lambdav^*) \\\end{bmatrix}$
for the exemplary dataset under $H_0$, and therefore $A=[1,  - I_{\beta\gammav}'(\beta_0,\gammav^*)  \, I_{\gammav\gammav}^{-1}(\beta_0,\gammav^*)]$.

As $n\rightarrow \infty$,  the null variance $n^{-1}V_n=n^{-1}[I_{\beta\beta_n}(\beta_0,\hat\lambdav) -I_{\beta\lambdav_n}'(\beta_0,\hat\lambdav) I_{\lambdav\lambdav_n}^{-1}(\beta_0,\hat\lambdav) I_{\beta\lambdav_n}(\beta_0,\hat\lambdav)] $ converges in probability to 
\begin{equation}\label{var0score} 
 \sigma_0^2 = I_{\beta\beta}(\beta_0,\lambdav^*) -I_{\beta\gammav}'(\beta_0,\lambdav^*) I_{\lambdav\lambdav}^{-1}(\beta_0,\lambdav^*) I_{\beta\lambdav}(\beta_0,\lambdav^*).
\end{equation}

\subsection{Power and Sample Size formulae}\label{sizesec}
The power of the score test \eqref{scoretest} is given by  
\begin{eqnarray}\label{power}
\begin{aligned}
P & = \Phi\left( \frac{\sqrt{n}|E_\beta|}{\sqrt{\sigma_1^2}}  - z_{1-\frac{\alpha}{2}} \sqrt{\frac{\sigma_0^2}{\sigma_1^2}}\,\right),
\end{aligned}
\end{eqnarray}
where $\Phi(\cdot)$ is the standard normal cumulative distribution function, and $\sigma_0^2$ and $\sigma_1^2$ are defined in equations \eqref{distscore}  and    \eqref{var0score}.
 Inverting \eqref{power} yields the sample size
\begin{equation}\label{size}
N_{new}= \frac{(z_{1-\alpha/2} \sigma_0+z_{P} \sigma_1)^2}{E_\beta^2}.
\end{equation}

\cite{self:1988} method is  formulated on basis of the noncentral chi-squared distribution, and the power and sample size estimates 
can be well approximated by
\begin{eqnarray}\label{size3}
\begin{aligned}
P_{SM}  & = \Phi\left( \frac{\sqrt{n}|E_\beta|}{\sqrt{\sigma_1^2}}  - z_{1-\frac{\alpha}{2}} \,\right), \\
N_{SM} & = \frac{(z_{1-\alpha/2} +z_{P} )^2\sigma_{1}^2}{E_\beta^2}.
\end{aligned}
\end{eqnarray}
It assumes $\sigma_0^2\approx \sigma_1^2$. The assumption holds under a sequence of contiguous alternatives.
 \cite{self:1992} showed that the performance of  the \cite{self:1988} procedure may degrade when the effect size is large or when the group sample sizes are unbalanced.

 It is generally easier to compute $\sigma_0^2$ than $\sigma_1^2$. Under  contiguous alternatives,  the power and sample size can also be calculated as
\begin{eqnarray}\label{size1}
\begin{aligned}
P_{s0}  & = \Phi\left( \frac{\sqrt{n}|E_\beta|}{\sqrt{\sigma_0^2}}  - z_{1-\frac{\alpha}{2}} \,\right), \\
N_{s0} &= \frac{(z_{1-\alpha/2} +z_{P} )^2\sigma_0^2}{E_\beta^2}.
\end{aligned}
\end{eqnarray}

  \begin{table}[htb]
\begin{center}
{\footnotesize
\begin{tabular}{c@{\extracolsep{5pt}}c@{\extracolsep{5pt}}c@{\extracolsep{5pt}}c@{\extracolsep{5pt}}c@{\extracolsep{5pt}}c@{\extracolsep{5pt}}c@{\extracolsep{5pt}}c@{\extracolsep{5pt}}c@{\extracolsep{5pt}}c@{\extracolsep{5pt}}c@{\extracolsep{5pt}}c@{\extracolsep{5pt}}c@{\extracolsep{5pt}}c@{\extracolsep{5pt}}cc} \\\hline  
dropout   & event  && & &&&  & & \multicolumn{5}{c}{power ($\%$) at $N_{new}$} \\ \cline{10-14}
proportion  & rate &&& \multicolumn{5}{c}{total sample size estimates} &&  \multicolumn{4}{c}{nominal power}\\ \cline{5-9}\cline{11-14}
$w_c \,(\%)$ & $\lambda_0$   & $\kappa$ & $\tau_c$ & $N_{new}$ & $N_{SM}$ & $N_{s0}$  &   ZL$^{(a)}$ & Wald$^{(b)}$ & SIM$^{(c)}$  & $P_{new}$ & $P_{SM}$  & $P_{s0}$ &   ZL$^{(a)}$  \\\hline 
\multicolumn{14}{c}{target power $80\%$}\\

      0&1.1& 0.9&  3&   58&   38&   68&   51&   54&80.85&80.29&93.83&73.59&84.74\\
             0&1.1& 1.2&  3&   70&   46&   82&   63&   65&80.80&80.48&93.57&73.96&84.38\\
             0&0.8& 0.9&  3&   65&   44&   76&   58&   61&80.91&80.10&92.87&73.96&84.52\\
             0&0.8& 1.2&  3&   77&   52&   89&   69&   73&80.84&80.24&92.77&74.17&84.22\\
             0&1.1& 0.9&  1&   96&   72&  107&   86&   94&81.05&80.22&89.94&75.71&84.20\\
             0&1.1& 1.2&  1&  108&   81&  121&   97&  105&80.88&80.17&90.22&75.51&84.02\\
             0&0.8& 0.9&  1&  117&   93&  128&  105&  116&81.15&80.17&88.48&76.38&83.95\\
             0&0.8& 1.2&  1&  129&  101&  143&  117&  127&81.02&80.09&88.87&76.09&83.82\\
            25&1.1& 0.9&  3&   64&   42&   74&   54&   59&81.07&80.60&93.42&74.21&86.36\\
            25&1.1& 1.2&  3&   76&   51&   88&   65&   71&80.62&80.38&93.04&74.17&85.77\\
            25&0.8& 0.9&  3&   72&   50&   83&   61&   68&80.93&80.20&92.33&74.39&85.92\\
            25&0.8& 1.2&  3&   85&   58&   97&   73&   80&81.03&80.51&92.39&74.72&85.85\\
            25&1.1& 0.9&  1&  108&   83&  120&   94&  105&81.12&80.26&89.43&76.03&85.22\\
            25&1.1& 1.2&  1&  121&   92&  135&  105&  117&81.12&80.28&89.76&75.87&85.24\\
            25&0.8& 0.9&  1&  132&  106&  144&  116&  131&81.10&80.11&87.95&76.56&84.73\\
            25&0.8& 1.2&  1&  145&  115&  159&  127&  143&80.87&80.07&88.38&76.31&84.79\\

\multicolumn{14}{c}{target power $90\%$}\\
         
           0&1.1& 0.9&  3&   74&   50&   91&   69&   72&89.88&90.41&97.69&83.30&91.97\\
             0&1.1& 1.2&  3&   89&   61&  109&   84&   87&89.75&90.37&97.52&83.50&91.65\\
             0&0.8& 0.9&  3&   83&   59&  101&   78&   82&89.90&90.16&97.20&83.65&91.81\\
             0&0.8& 1.2&  3&   98&   70&  119&   93&   97&89.60&90.12&97.11&83.72&91.54\\
             0&1.1& 0.9&  1&  124&   97&  143&  116&  125&90.25&90.17&95.73&85.53&91.79\\
             0&1.1& 1.2&  1&  139&  108&  162&  131&  140&89.93&90.07&95.83&85.24&91.60\\
             0&0.8& 0.9&  1&  152&  124&  172&  143&  155&90.50&90.13&94.96&86.28&91.72\\
             0&0.8& 1.2&  1&  167&  135&  191&  158&  170&90.24&90.02&95.13&85.92&91.57\\
            25&1.1& 0.9&  3&   81&   56&   99&   73&   79&89.88&90.26&97.39&83.54&92.91\\
            25&1.1& 1.2&  3&   97&   68&  118&   88&   95&89.80&90.34&97.29&83.82&92.72\\
            25&0.8& 0.9&  3&   92&   66&  111&   83&   91&89.99&90.16&96.92&84.05&92.80\\
            25&0.8& 1.2&  3&  108&   78&  130&   98&  107&89.73&90.18&96.88&84.13&92.64\\
            25&1.1& 0.9&  1&  139&  111&  160&  127&  141&90.27&90.02&95.36&85.68&92.41\\
            25&1.1& 1.2&  1&  156&  123&  180&  142&  157&90.15&90.13&95.59&85.61&92.49\\
            25&0.8& 0.9&  1&  172&  142&  193&  157&  175&90.48&90.13&94.69&86.53&92.33\\
            25&0.8& 1.2&  1&  189&  154&  213&  172&  192&90.24&90.17&94.96&86.34&92.40\\

\hline
 \end{tabular} \caption{Estimated sample size at the target $ 90\%$ power and estimated power for the score test from NB regression in superiority trials \newline
$^{(a)}$ Method $3$ of  \cite{zhu:2014}. It  estimates the null variance of the test statistic based on the approximate restricted MLE.\newline
$^{(b)}$ Sample size estimate by  \cite{tang:2015} method for  Wald test is displayed  for comparison \newline
$^{(c)}$ Simulated power (SIM) are evaluated at $N_{new}$ based on $160,000$ simulated datasets.
}\label{power_sim1}
}
\end{center}

\end{table}

\section{Sample size  for  NB regression}\label{nbscore}

The NB regression has been widely used to analyze overdispersed count data and recurrent event data. The NB distribution can be written as a Poisson–gamma mixture. If $Y$ follows a Poisson
distribution with mean $\epsilon \mu$, where $\epsilon$ is gamma distributed with mean $1$ and variance $\kappa$, 
the marginal distribution of Y is $NB(\mu,\kappa)$
\begin{equation}\label{nbdist}
\Pr(Y=y|\mu,\kappa) = \frac{\Gamma(y+1/\kappa)}{y! \Gamma(1/\kappa)} \left[ \frac{\kappa \mu}{1+\kappa\mu} \right]^y \left[ \frac{1}{1+\kappa\mu} \right]^{1/\kappa}, \,\,\, y=0,1,2,\ldots,
\end{equation}
where $\Gamma(\cdot)$ is the Gamma function.

Suppose in a trial, $n$ subjects are assigned randomly to either the experimental
$(g=1)$ or  control $(g=0)$ treatment group. Let $n_g$ be the number of
subjects in group $g$. 
We assume the planned treatment duration  is $\tau_c$ for each subject, but subjects may discontinue the study with the loss-to-follow-up distribution $G(t)$.
Let $t_i$ be the follow-up time, and $y_i$ the number of events for subject $i$.
Then $y_i|g_i=g,t_i \sim NB(\lambda_g t_i, \kappa)$, where  $\lambda_g$ is the event rate  in group $g$.

Suppose a lower event rate indicates better health status. 
In a superiority trial, the purpose is to demonstrate that  the experimental treatment can reduce the event rate relative to the control treatment.
  The hypothesis can be written as
\begin{equation}\label{hypothesis}
   H_0: \frac{\lambda_1}{\lambda_0}=1 \text{ \it vs }  H_1:   \frac{\lambda_1}{\lambda_0}<1.
\end{equation}

 In a NI trial, the objective  \citep{tang2:2017, tang:2018} is to show that the test treatment is not materially less efficacious than a standard control treatment  by proving $\lambda_1/\lambda_0<M_0$, 
where $M_0$ is the prespecified margin that is bigger than $1$, but close to $1$. 
  The hypothesis can be written as
\begin{equation}\label{hypothesis}
   H_0: \frac{\lambda_1}{\lambda_0}=M_0 \text{ \it vs }  H_1:   \frac{\lambda_1}{\lambda_0}<M_0.
\end{equation}
Mathematically, the superiority trial  can be viewed as a special case of the NI trial by setting  $M_0=1$.  

Let $\beta=\log(\lambda_1/\lambda_0)$, $\beta_0=\log(M_0)$ and $\alpha=\log(\lambda_0)$. Since $\mu_i = \alpha + \log(t_i)+ g_i\log(M_0)$ under $H_0$,
the null model can be easily fitted using standard software packages (e.g. SAS Proc Genmod) by setting the offset as $\log(t_{i})$ for subjects in the control group,
and $\log(M_0 t_{i})$ for subjects in the experimental arm.
The score test can be written as 
\begin{equation}\label{score}
 Z_s  =\frac{\sum_{\{i: g_i=1\}} \frac{ y_{i}-\hat\mu_{i}}{1+\hat\kappa\hat\mu_{i}}}{ \sqrt{\frac{\hat{d}_0\hat{d}_1}{\hat{d}_0+\hat{d}_1}}},
\end{equation}
where $\hat\mu_{i}=\exp(\hat\alpha+\beta_0 g_i) t_{i}$,  and   $\hat{d}_g= \sum_{\{i: g_i=g\}} \hat{\mu}_{i}/(1+\hat\kappa \hat\mu_{i})$ for $g=0$ and $1$.

The power and sample size  can be calculated using the procedure described in Section \ref{scoreglm}.
The expressions for the score function, observed and expected information matrix are given  in equations (2.3)-(2.8) in  \cite{lawless:1987}.
 In our implementation, the continuous time to follow-up is approximated  by a categorical variable with $L=100$ levels 
\begin{eqnarray}\label{discrete}
t = \begin{cases}
t_{L} =  \tau_c &    \text{ with probability  $p_{L}=1 - G(\tau_c)$ }\\
t_{l}= G^{-1}\left( \frac{ p_L (l-0.5)}{L-1}\right) & \text{ with probability  $(1-p_L)/(L-1)$  for  $l=1, \ldots,L-1$}. 
\end{cases}
\end{eqnarray}
The final result is  insensitive to the choice of $L$ if $L$ is not too small.
We allow the loss-to-follow-up distribution to differ by the treatment group. There are $m=2L$ possible combinations of the values for the treatment and time to follow-up.
We truncate the number of response categories at a large number $J=200$ so that $\Pr(y_{ij} \geq J)< 10^{-5}$.
The  full exemplary dataset consists of $mJ$ observations. 
The weights for the $mJ=2LJ$ observations are calculated according to the treatment allocation ratio and the true distribution
defined in equation \eqref{nbdist}. It requires the specification of the dispersion parameter $\kappa$ and the event rates for each group. As mentioned in Section \ref{asyscore},
we  {\color{red} multiply} all the weights by a large value (say $10^9$) to minimize the effect of truncation since the SAS Genmod procedure truncates the weight to an integer.  

When all subjects have equal follow-up time ($t_{gj}\equiv t$), the method can be slightly simplified with $m=2$. In this paper,
we focus on the analysis of recurrent events. 
The simplified procedure is also  suitable for other types of overdispersed counts  such as the number of magnetic resonance imaging lesions in multiple sclerosis trials.
Let $\mu_g$ and $\bar{y}_g$ be, respectively, the expected and observed mean count in group $g$. For recurrent events, $\mu_g=\lambda_g t$. 
 The score test \eqref{score} reduces to
\begin{equation}\label{score2}
 W_s  =\frac{\bar{y}_1 -M_0 \bar{y}_0}{ \sqrt{n_1^{-1}(\hat{\mu}_1+\hat\kappa\hat\mu_1^2)+ n_0^{-1}(\hat{\mu}_0+\hat\kappa\hat\mu_0^2)}},
\end{equation}
where $\hat\mu_1=M_0\hat\mu_0$ is the MLE under $H_0$. Test \eqref{score2} is  similar to
the test (equation (7)) of  \cite{farrington:1990} for assessing the relative risk between two binomial proportions. In this special case,
the sample size in the control arm is 
\begin{equation}\label{sizesimp}
n_0= \frac{\left[z_{1-\alpha/2} \sqrt{\theta^{-1} (\mu_1^*+\kappa^*\mu_1^{*^2})+ (\mu_0^*+\kappa^*\mu_0^{*^2})}  +z_{P} \sqrt{\theta^{-1} (\mu_1+\kappa\mu_1^2)+ (\mu_0+\kappa\mu_0^2)}\right]^2}{(\mu_1-M_0\mu_0)^2},
\end{equation}
where $\theta=n_1/n_0$ and  $(\mu_0^*,\mu_1^*,\kappa^*)$ can be estimated by the exemplary dataset method.  In superiority trials, we can approximate $(\mu_0^*,\mu_1^*,\kappa^*)$ by the method of moments,
\begin{equation}\label{nbzhusol}
\mu_0^*=\mu_1^*=\bar{\mu}=\frac{\theta \mu_1+\mu_0}{\theta+1} \text{ and }
         \kappa^*= \frac{ \kappa (\theta \mu_1^2 +\mu_0^2)}{(\theta+1)\bar\mu^2}+\frac{\theta(\mu_1-\mu_0)^2}{(\theta+1)^2\bar\mu^2},
\end{equation}
where $\kappa^*>\kappa$ if $\mu_1\neq \mu_0$, and $\kappa^*$ is the solution to 
$$ n_1 \frac{ \text{E}( y_1 -\mu_1^*)^2}{\mu_1^*(1+\kappa^*\mu_1^*)} +n_0\frac{ \text{E}( y_0 -\mu_0^*)^2}{\mu_0^*(1+\kappa^*\mu_0^*)} =n_1+n_0.$$

\begin{figure}
\centering
  \includegraphics[scale=0.30]{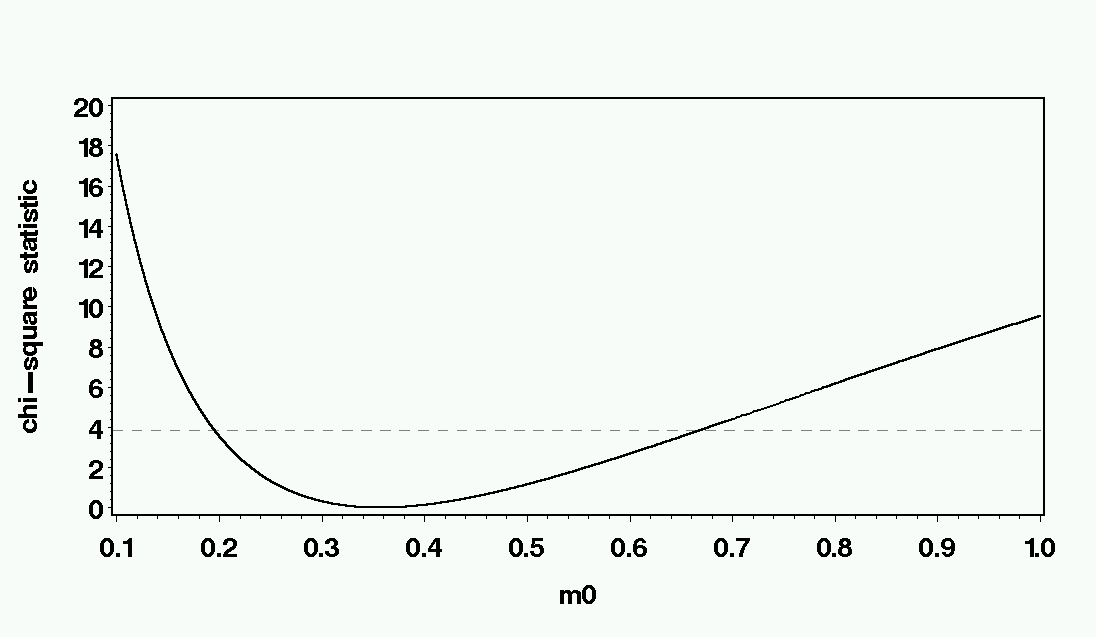}
\caption{Plot of the chi-square statistic ($Z^2(\beta)$ from the score test) as a function of $M_0$ in  a CGD trial}
\label{fig:cgi}
\end{figure}

It would be interesting to compare the proposed method with that recently developed by \cite{zhu:2014} and  \cite{zhu:2017} for superiority and NI trials because they use a similar idea  to the score test.  The approaches of
\cite{zhu:2014} and  \cite{zhu:2017} are based on  the statistic $\log(\hat\mu_1/\hat\mu_0)-\log(M_0)$ instead of the score statistic 
\begin{equation}\label{sizezhu}
n_0= \frac{\left[z_{1-\alpha/2} \sqrt{\frac{1}{\mu_0^{**}}+\frac{1}{\theta \mu_1^{**}}+ \frac{1+\theta}{\theta}\kappa}  +z_{P} \sqrt{\frac{1}{\mu_0}+\frac{1}{\theta \mu_1}+ \frac{1+\theta}{\theta}\kappa}\,\right]^2}{\left[\log(\mu_1/\mu_0)-\log(M_0)\right]^2},
\end{equation}
where $\kappa$ is assumed to be known, 
 $a =-\kappa M_0 (1+\theta)$, $b=\kappa (\mu_0 M_0+ \theta \mu_1)-(1+\theta M_0)$, $c=\mu_0+\theta\mu_1$, and
$\mu_0^{**}=([-b-\sqrt{b^2-4ac}]/2a$ and $\mu_1^{**}=M_0\mu_0^{**}$ are the limiting values of the restricted MLE at given $\kappa$.
\cite{zhu:2014} and  \cite{zhu:2017}  implicitly make two approximations. Firstly, the follow-up time  is set to their mean values (i.e. $t_{i}=\bar{t}$) for all individuals, leading to 
underestimated variance of  $\log(\hat\mu_1/\hat\mu_0)-\log(M_0)$ under both $H_0$ and $H_1$ (this can be proved by using the inequality 
in Appendix A.2 of  \cite{tang:2015}). 
 Secondly, it approximates $\kappa^*$ by $\kappa$, and  the null variance of  $\log(\hat\mu_1/\hat\mu_0)-\log(M_0)$ is usually underestimated 
since  $\kappa^*$  obtained under the null hypothesis in the score approach tends to be larger than $\kappa$ particularly when the treatment effect is large. 
This is shown in equation \eqref{nbzhusol} for superiority trials when all subjects have equal follow-up time.
The phenomenon is analogous to the comparison of 
two groups with continuous outcomes, in which the variance estimate based on the pooled outcomes $y_{i0}\sim N(\mu_0,\sigma^2)$ and $y_{i1}\sim N(\mu_1,\sigma^2)$ tends to  overestimate the true variance 
if the mean difference is ignored.
Therefore  Zhu-Lakkis's approach tends to  underestimate the sample size.
In superiority trials ($M_0=1$) with equal treatment allocation,  Zhu-Lakkis's sample size estimate is strictly smaller
than the lower sample size bound of  \cite{tang:2015} for the Wald test from the NB regression  \citep{tang2:2017}.

Below we present several examples to illustrate the proposed method.

\begin{example}\label{exam1}
{\normalfont
  Chronic granulomatous disease (CGD) is a rare inherited disorder of  the  immune system, characterized by recurrent pyogenic infections.
Suppose we plan to design a   two-arm  CGD  trial to assess the effect of an experimental treatment on the infection rate. Some parameters are estimated from a CGD trial
  analyzed by \cite{matsui:2005} and \cite{tang:2018e}. The historical trial enrolled $n=128$ eligible patients.
It was terminated early for efficacy based on an interim analysis. In the trial,  $14$ ($22.2\%$) out of $63$ patients in the gamma interferon group and $30$ ($46.2\%$) out of 65 patients on placebo had at least one serious infection. 
We analyze the number of repeated infections using the NB regression.
The event rate ratio between two treatments based on the Wald statistic is $0.3566$ ($95\%$ CI: $[0.1934, 0.6575]$).  
Figure \ref{fig:cgi} plots the chi-square statistic (i.e. $Z^2(\beta)$ from the score test) as a function of $M_0$. The score CI  is $[0.1957,0.6681]$, which corresponds to the region $\{\beta: Z^2(\beta) \leq z_{1-\alpha/2}^2=3.814\}$.
The score CI is slightly wider than the Wald CI.

We estimate the sample size at the following parameter values. The infection rate is  $\lambda_0=1.1$ infections per year in the control arm, and $\kappa=0.9$, which are close to the unconstrained MLE ($\hat\lambda_0=1.07$, $\hat\kappa=0.91$) from the analysis of the historical CGD trial.
Suppose the experimental treatment can reduce the infection rate by $60\%$ (i.e. $\lambda_1/\lambda_0=0.4$). The target power is $80\%$ or $90\%$, and the  two-sided significance level is $\alpha= 0.05$.
The treatment allocation ratio is $1:1$.
The planned treatment duration is $\tau_c=1$ or $3$ years for each subject, but subjects may discontinue the trial early with a $w_c=25\%$ chance and the loss to follow-up is  exponentially distributed. 
We also assess the performance of the proposed method  at other parameter values ($\lambda_0=0.8$, $\kappa=1.2$, $w_c=0$).

We compare several sample size procedures for the NB regression. In  \cite{zhu:2014}, three methods were proposed to evaluate the variance of the test statistic under $H_0$. 
We evaluate only the approach recommended by the authors, in which the null variance is calculated based on the approximate restricted MLE.
The sample size estimate by \cite{tang:2015} method for the Wald test is also presented for the purpose of comparison. 
Table \ref{power_sim1} reports  sample size estimates at the target  power, and the power estimates at the  sample size $N_{new}$ determined by equation \eqref{size}.  
The simulated  power is evaluated based on $160,000$ trials. There is more than $95\%$ chance
that the simulated power lies within $2 \sqrt{ 0.8*0.2/160000}=0.2\%$ and $2 \sqrt{ 0.9*0.1/160000}=0.15\%$ of the true power respectively when the target power is $80\%$ and $90\%$.
 Because the estimated sample size is rounded up to the nearest integer, the nominal power by the proposed method at $N_{new}$  is slightly larger than the target power.
The proposed method yields the  nominal power estimate that is generally within $1\%$ of the simulated power, evidencing the accuracy of the proposed method. 

Because the treatment effect is quite large ($\lambda_1/\lambda_0=0.4$), 
the variances of the score statistic under $H_0$ and $H_1$ are not close. 
  Self and Mauritsen's method  (equation \eqref{size3})  underestimates the required size while formula \eqref{size1} overestimates the sample size.
 \cite{tang2:2017, tang:2018} demonstrates that the method of  \cite{zhu:2014}  underestimates the required size
if the follow-up time varies across patients. Table \ref{power_sim1} indicates that Zhu-Lakkis's formula still underestimates the sample size and overestimates the power
even if all patients have equal follow-up time.   
At the target $90\%$ power,  the sample size estimates are quite close for the Wald  and score tests.
 But at the target $80\%$ power, the score test requires  about $7.0\%$ to $8.5\%$ more subjects than the Wald test when $\lambda_0=1.1$, $\tau_c=3$ and $w_c=0$ or $25\%$ in this example.}

\end{example}

\begin{example}\label{exam3}
{\normalfont
We compare the proposed method,   Self and Mauritsen's procedure and \cite{zhu:2017}  approach  for the sample size determination in NI trials. 
For the purpose of illustration, we set $\lambda_0=1.0$ or $1.5$, $w_c=0$ or $25\%$, $\kappa=1$, $\lambda_1/\lambda_0=0.8$ or $1$, and $M_0=1.25$.
In practice, one typically assumes the true rate ratio is $1$ or close to $1$ in designing a NI trial. 
 If the experimental treatment is truly more effective than the control treatment, but the treatment effect is relatively small (e.g. true rate ratio $=0.8$), a NI trial may be chosen
if it is infeasible to run a  superiority trial that requires a much larger sample size. 
If the NI test is significant, one may continue to conduct a superiority test to assess whether the experimental treatment is more effective than the control treatment. 
In this example, we focus  on determining the sample size and power for the NI test.
The simulated  power is evaluated based on $160,000$ trials. There is more than $95\%$ chance
that the simulated power lies within $2 \sqrt{ 0.8*0.2/160000}=0.2\%$ of the true power.

Table \ref{power_sim2} reports  sample size estimates at the target $P=80\%$ and $90\%$ power and $\alpha=0.05$. Since the NI test is one-sided, the actual 
type I error rate is $\alpha/2=0.025$.  Because the treatment effect is smaller in NI trials than in superiority trials, the differences in the sample size estimates from various methods reduce  in NI trials, but the general pattern is similar to that in Example \ref{exam1}. 
}
\end{example}

  \begin{table}[h]
\begin{center}{
\begin{tabular}{c@{\extracolsep{5pt}}c@{\extracolsep{5pt}}c@{\extracolsep{5pt}}c@{\extracolsep{5pt}}c@{\extracolsep{5pt}}c@{\extracolsep{5pt}}c@{\extracolsep{5pt}}c@{\extracolsep{5pt}}c@{\extracolsep{5pt}}c@{\extracolsep{5pt}}c@{\extracolsep{5pt}}c@{\extracolsep{5pt}}c@{\extracolsep{5pt}}c@{\extracolsep{5pt}}c@{\extracolsep{5pt}}c@{\extracolsep{5pt}}c@{\extracolsep{5pt}}c@{\extracolsep{5pt}}c} \\\hline  
dropout                          & event&true& & &&&&  & & \multicolumn{5}{c}{power ($\%$) at $N_{new}$} \\ \cline{10-14}
proportion                         &rate & rate & &\multicolumn{5}{c}{total sample size estimates} && \multicolumn{4}{c}{nominal power}  \\\cline{5-9} \cline{11-14}
$w_c\, (\%)$ & $\lambda_0$  & ratio & $\kappa$ & $N_{new}$ & $N_{SM}$ & $N_{s0}$    & Zhu$^{(a)}$ & Wald$^{(b)}$ &  SIM$^{(c)}$  & $P_{new}$ & $P_{SM}$  & $P_{s0}$ &   Zhu$^{(a)}$ \\\hline 
    
             0&1.0& 0.8&  1&  337&  313&  348&  334&  335&79.93&80.02&82.92&78.76&80.41\\
             0&1.5& 0.8&  1&  278&  253&  290&  275&  276&79.88&80.02&83.63&78.45&80.47\\
            25&1.0& 0.8&  1&  377&  351&  388&  361&  375&80.15&80.10&82.77&78.93&81.76\\
            25&1.5& 0.8&  1&  308&  282&  319&  293&  306&80.09&80.09&83.46&78.62&82.00\\
             0&1.0& 1.0&  1& 1264& 1244& 1272& 1264& 1262&80.01&80.02&80.62&79.76&80.01\\
             0&1.5& 1.0&  1& 1053& 1031& 1063& 1053& 1051&80.07&80.01&80.84&79.65&80.02\\
            25&1.0& 1.0&  1& 1407& 1388& 1416& 1360& 1405&79.85&80.00&80.55&79.77&81.33\\
            25&1.5& 1.0&  1& 1161& 1139& 1171& 1117& 1160&79.95&80.01&80.78&79.68&81.53\\
\hline
 \end{tabular} \caption{Estimated sample size at the target $ 80\%$ power and estimated power for the score test from NB regression in    NI trials\newline
$^{(a)}$ Method $3$ of   \cite{zhu:2017}. It  estimates the null variance of the test statistic based on the  approximate restricted MLE.\newline
$^{(b)}$ Sample size estimate by  \cite{tang:2015} method for  Wald test is displayed  for comparison \newline
$^{(c)}$ Simulated power (SIM) are evaluated at $N_{new}$ based on $160,000$ simulated datasets.
}\label{power_sim2}
}
\end{center}
\end{table}

\section{Sample size   for  logistic regression}\label{logissec}
 \cite{self:1988} investigated the sample size estimation for comparing two binomial proportions ($g_i=0$ or $1$) using logistic regression while controlling for a categorical covariate $z_i$ with $S=2$ levels.
We call $z_i$ a stratum variable, and revisit the problem with $S\geq 2$ strata.   
 Suppose  for subjects in stratum $s$,  $y_{i}$ follows a Bernoulli distribution with the probability of success
\begin{equation}\label{logismod}
 \Pr (y_i=1 | g_i, z_i=s) =  \frac{\exp(\alpha_0 +\alpha_s + \beta g_i)}{1+\exp(\alpha_0 +\alpha_s + \beta g_i)}= \frac{1}{1+\exp(-\alpha_0 -\alpha_s - \beta g_i)},
\end{equation}
where  $\exp(\alpha_0)$ is the  odds at $g_i=0$ in stratum $1$,
$\exp(\beta)$ is the odds ratio associated with the group status $g_i$ among subjects from the same stratum, and $\exp(\alpha_s)$ is the odds ratio for subjects in stratum $s$ relative to subjects with the same $g_i$ from stratum $1$ ($\alpha_1=0$).
 
Model \eqref{logismod} can be used to analyze data from both perspective clinical trials and retrospective case-control studies. The objectives are different, but the underlying statistical problems are similar
in the two types of studies.  Table \ref{table2strata} displays the data format for both studies. 
In a clinical trial, we compare the proportion of responders between two treatment groups, where
$x_{gs}$ is the number of responders  among $n_{gs}$ subjects assigned to   treatment group $g$ in stratum $s$.
 In the case-control study, the aim is to compare the  proportion of exposed between the case and control groups, where
 $x_{gs}$ is the number of exposed subjects among $n_{gs}$  case ($g=1$) or control ($g=0$) subjects  in stratum $s$.
 \begin{table}[h]
\centering
{\footnotesize
\begin{tabular}{l@{\extracolsep{5pt}}c@{\extracolsep{5pt}}c@{\extracolsep{5pt}}c@{\extracolsep{5pt}}c@{\extracolsep{5pt}}c@{\extracolsep{5pt}}c@{\extracolsep{5pt}}c@{\extracolsep{5pt}}c@{\extracolsep{5pt}}ccc}\\\hline 
  \multicolumn{6}{c}{Clinical trials }  &   \multicolumn{6}{c}{Case-control study } \\\cline{1-6}\cline{7-12}
      & \multicolumn{2}{c}{ $z=1$} &  & \multicolumn{2}{c}{ $z=S$} & & \multicolumn{2}{c}{ $z=1$} & & \multicolumn{2}{c}{ $z=S$} \\\cline{2-3}\cline{5-6} \cline{8-9}\cline{11-12}
    & placebo  & active  & $\ldots$& placebo   & active  &    exposure &control  & case & $\ldots$ & control   & case  \\
 Event              &$g=0$  &$g=1$      &$\ldots$& $g=0$  &$g=1$      & status&$g=0$  &$g=1$      &$\ldots$& $g=0$  &$g=1$\\ \hline
Yes  &   $x_{10}$ & $x_{11}$  &$\ldots$&  $x_{S0}$ & $x_{S1}$  & exposed &   $x_{10}$ & $x_{11}$  &$\ldots$&  $x_{S0}$ & $x_{S1}$\\
No  & $n_{10} -x_{10}$ & $n_{11} -x_{11}$  &$\ldots$&  $n_{S0} -x_{S0}$ & $n_{S1} -x_{S1}$ & No & $n_{10} -x_{10}$ & $n_{11} -x_{11}$  &$\ldots$&  $n_{S0} -x_{S0}$ & $n_{S1} -x_{S1}$ \\
Sum &        $n_{10}$ & $n_{11}$   &$\ldots$&$n_{S0}$ & $n_{S1}$    & Sum &        $n_{10}$ & $n_{11}$   &$\ldots$&$n_{S0}$ & $n_{S1}$ \\            \hline   
\end{tabular}
}\caption{Binary outcomes from stratified clinical trials and case-control studies  }\label{table2strata}
\end{table}

In fact, the score test for testing $H_0:\beta=0$ has explicit analytic expression 
\begin{equation}\label{supor}
 Z = \frac{\sum_{s=1}^S \frac{n_{s1}n_{s0}}{n_s}( \hat{p}_{s1} -\hat{p}_{s0}) }{\sqrt{\sum_{s=1}^S \frac{n_{s1}n_{s0}}{n_s} \bar{p}_s(1-\bar{p}_s)}},
\end{equation}
where $\hat{p}_{sg}=x_{sg}/n_{sg}$, $n_s=n_{s0}+n_{s1}$, and $ \bar{p}_s= (x_{s0}+x_{s1})/n_s$.
 The power and sample size formulae in Section \ref{sizesec} can be used by setting
\begin{eqnarray}\label{logisvar}
\begin{aligned}
E_\beta &=\sum_{s=1}^S  t_s\rho_s(1-\rho_s) ( p_{s1} - p_{s0}),\\
 \sigma_0^2 &= \sum_{s=1}^S t_s\rho_s(1-\rho_s) p_s^*(1-p_s^*), \\
\sigma_1^2&=  \sum_{s=1}^S  t_s\rho_s(1-\rho_s) [(1-\rho_s)p_{s1}(1-p_{s1})+ \rho_s p_{s0}(1-p_{s0})],
\end{aligned}
\end{eqnarray}
where $p_{sg}=\text{E}(\hat{p}_{sg})$ is the true response rate, $n=\sum_{s=1}^S n_s$ is the total sample size, 
$t_s=\frac{ n_s}{n}$ is the proportion of subjects contributed by stratum $s$, $\rho_s=\frac{n_{s1}}{n_s}$ is the proportion of subjects
from group $g=1$ in stratum $s$,  and $p^*_s= \rho_s p_{s1}+(1-\rho_s)p_{s0}$. 
The technical details are omitted here. We will extend the method to  sample size determination for the stratified score tests in superiority, NI and equivalence trials  on basis of the risk difference, relative risk or odds ratio effect measures in \cite{tang:2019b}, and a general proof will be presented in that paper.
The score statistic \eqref{supor} is identical to \cite{cochran:1954} statistic, and the power formula \eqref{power} is identical to that derived by \cite{nam:1992} for Cochran's test although  \cite{nam:1992} considered only the case-control studies.

We conduct two simulation studies to compare several methods. 

\begin{example}
\normalfont
Suppose there are two strata. Let $(\Pi_1, \Pi_2,\Pi_3,\Pi_4)$ denote respectively the proportion of subjects  with $(g_i, z_i)=(0,1)$, $(0,2)$, $(1,1)$ and $(1,2)$. Thus $t_1=\Pi_1+\Pi_3$, $t_2=\Pi_2+\Pi_4$, $\rho_1=\frac{\Pi_3}{t_1}$ and $\rho_2=\frac{\Pi_4}{t_2}$.
We set $(\Pi_1,\Pi_2,\Pi_3,\Pi_4)=(.25,.25,.25,.25)$ and $(.4,.1,.1,.4)$. The odds ratio associated with the stratum  is $\exp(\alpha_1)=2$, and the odds ratio for the exposure is  $\exp(\beta)=2$ or $3$.
The overall response rate in the study population 
$$ \bar\mu=   \frac{\Pi_1}{1+\exp(-\alpha_0 )} +  \frac{ \Pi_2}{1+\exp(-\alpha_0 -\alpha_1  )} +     \frac{ \Pi_3 }{1+\exp(-\alpha_0  - \beta )}     +\frac{\Pi_4}{1+\exp(-\alpha_0 -\alpha_1- \beta )}$$
is set to $0.15$ and $0.5$, which is used to derive $\alpha_0$. The set up is similar to that reported in Table 2 of \cite{self:1988}.

Table \ref{power_logis1} displays the power and sample size results at the target power $80\%$, $90\%$ and $95\%$, and two-sided type I error $0.05$.
The analytic expression \eqref{size3} gives the same sample size estimates as that reported in table 2 of \cite{self:1988} in all cases at $\bar\mu=0.5$, but slightly larger estimates in all cases at $\bar\mu=0.15$ possibly due to rounding errors.
This verifies the validity of the power and sample size calculation based on the simpler equation \eqref{logisvar}   in the logsitic regression. 
The simulated power is estimated at the sample size $N_{new}$ from the proposed method based on $10^6$ simulated datasets.
There is more than $95\%$ chance
that the simulated power lies within $2 \sqrt{ 0.8*0.2/10^6}=0.08\%$ of the true power.

All methods perform well possibly because the sample sizes are balanced overall between two groups ($\Pr(g=1)=\Pr(g=0)=0.5$) although when  $(\Pi_1,\Pi_2,\Pi_3,\Pi_4)=(.4,.1,.1,.4)$, the sample sizes are highly unbalanced between two groups within each stratum.
We compare the methods by assessing how close the estimated nominal power is to the empirical power at a given sample size.
There are more  cases with $>1\%$ difference between the nominal and simulated power estimates  by  formulae \eqref{size3} and \eqref{size1} than by formula \eqref{power}.
In nearly all cases at  $(\Pi_1,\Pi_2,\Pi_3,\Pi_4)=(.4,.1,.1,.4)$, the nominal power by formula \eqref{power} is closer to the simulated power than that by equation \eqref{size3}.

\end{example}

\begin{landscape}
  \begin{table}[h]
\begin{center}{
\footnotesize
\begin{tabular}{c@{\extracolsep{5pt}}c@{\extracolsep{5pt}}c@{\extracolsep{5pt}}c@{\extracolsep{5pt}}c@{\extracolsep{5pt}}c@{\extracolsep{5pt}}c@{\extracolsep{5pt}}c@{\extracolsep{5pt}}c@{\extracolsep{5pt}}c@{\extracolsep{5pt}}c@{\extracolsep{5pt}}c@{\extracolsep{5pt}}c@{\extracolsep{5pt}}c@{\extracolsep{5pt}}c@{\extracolsep{5pt}}c@{\extracolsep{5pt}}c@{\extracolsep{5pt}}c@{\extracolsep{5pt}}c@{\extracolsep{5pt}}c@{\extracolsep{5pt}}c@{\extracolsep{5pt}}c@{\extracolsep{5pt}}c@{\extracolsep{5pt}}c@{\extracolsep{5pt}}rcccccccccccccccccccccccccc} \\\hline  
                   &                    &  & \multicolumn{8}{c}{$(\Pi_1,\Pi_2,\Pi_3,\Pi_4)=(.25,.25,.25,.25)$} &\multicolumn{8}{c}{$(\Pi_1,\Pi_2,\Pi_3,\Pi_4)=(.4,.1,.1,.4)$}  \\\cline{4-11} \cline{12-19} 
                   &                     &target & & \multicolumn{3}{c}{estimated size} & & \multicolumn{3}{c}{ nominal power ($\%$)} & & \multicolumn{3}{c}{estimated size} & & \multicolumn{3}{c}{ nominal power ($\%$)} \\\cline{5-7}\cline{9-11}\cline{13-15}\cline{17-19}
$\bar\mu$ & $\exp(\beta)$   & power ($\%$) & $\alpha_0$ & $N_{new}$ & $N_{sw}$ & $N_{s0}$   &  SIM  & $P_{new}$ & $P_{SM}$  & $P_{s0}$   & $\alpha_0$ & $N_{new}$ & $N_{sw}$ & $N_{s0}$   &  SIM  & $P_{new}$ & $P_{SM}$  & $P_{s0}$              \\\hline 
                      
                                               $.15$&$ 2$&$ 80$&$ -2.5102$&$ 543$&$ 537$&$ 546$&$ 80.58$&$ 80.02$&$ 80.44$&$ 79.84$&$ -2.5597$&$ 882$&$ 853$&$ 894$&$ 80.51$&$ 80.04$&$ 81.32$&$ 79.49$\\
                                                &&$ 90$&$ -2.5102$&$ 726$&$ 719$&$ 730$&$ 90.37$&$ 90.02$&$ 90.28$&$ 89.84$&$ -2.5597$&$1175$&$1142$&$1197$&$ 90.25$&$ 90.02$&$ 90.81$&$ 89.49$\\
                                                &&$ 95$&$ -2.5102$&$ 897$&$ 890$&$ 903$&$ 95.25$&$ 95.01$&$ 95.16$&$ 94.88$&$ -2.5597$&$1449$&$1412$&$1480$&$ 95.13$&$ 95.01$&$ 95.47$&$ 94.61$\\
                                                &$ 3$&$ 80$&$ -2.7737$&$ 231$&$ 225$&$ 233$&$ 81.55$&$ 80.13$&$ 81.12$&$ 79.70$&$ -2.8507$&$ 382$&$ 360$&$ 392$&$ 81.24$&$ 80.05$&$ 82.32$&$ 79.07$\\
                                                &&$ 90$&$ -2.7737$&$ 308$&$ 301$&$ 312$&$ 90.98$&$ 90.08$&$ 90.69$&$ 89.67$&$ -2.8507$&$ 507$&$ 482$&$ 524$&$ 90.59$&$ 90.02$&$ 91.41$&$ 89.06$\\
                                                &&$ 95$&$ -2.7737$&$ 380$&$ 372$&$ 386$&$ 95.61$&$ 95.05$&$ 95.40$&$ 94.74$&$ -2.8507$&$ 624$&$ 596$&$ 648$&$ 95.42$&$ 95.02$&$ 95.82$&$ 94.28$\\
                                                $.50$&$ 2$&$ 80$&$ -0.6931$&$ 273$&$ 267$&$ 275$&$ 79.88$&$ 80.10$&$ 80.88$&$ 79.75$&$ -0.6931$&$ 417$&$ 432$&$ 411$&$ 80.05$&$ 80.04$&$ 78.63$&$ 80.65$\\
                                                &&$ 90$&$ -0.6931$&$ 364$&$ 358$&$ 368$&$ 89.88$&$ 90.03$&$ 90.52$&$ 89.71$&$ -0.6931$&$ 561$&$ 578$&$ 550$&$ 90.21$&$ 90.03$&$ 89.14$&$ 90.60$\\
                                                &&$ 95$&$ -0.6931$&$ 449$&$ 442$&$ 455$&$ 95.13$&$ 95.01$&$ 95.29$&$ 94.76$&$ -0.6931$&$ 696$&$ 715$&$ 679$&$ 95.00$&$ 95.02$&$ 94.49$&$ 95.45$\\
                                                &$ 3$&$ 80$&$ -0.8959$&$ 111$&$ 105$&$ 113$&$ 81.76$&$ 80.28$&$ 82.22$&$ 79.43$&$ -0.8959$&$ 166$&$ 173$&$ 163$&$ 80.15$&$ 80.10$&$ 78.42$&$ 80.82$\\
                                                &&$ 90$&$ -0.8959$&$ 147$&$ 141$&$ 151$&$ 89.63$&$ 90.09$&$ 91.29$&$ 89.27$&$ -0.8959$&$ 223$&$ 232$&$ 218$&$ 90.21$&$ 90.01$&$ 88.94$&$ 90.68$\\
                                                &&$ 95$&$ -0.8959$&$ 181$&$ 174$&$ 187$&$ 94.95$&$ 95.06$&$ 95.75$&$ 94.43$&$ -0.8959$&$ 277$&$ 286$&$ 270$&$ 94.94$&$ 95.02$&$ 94.39$&$ 95.52$\\

\hline
 \end{tabular} \caption{Power and sample size estimate for the score test from logistic regression with equal group sample sizes \newline
[1] Simulated power (SIM) are evaluated at $N_{new}$ based on $1000,000$ simulated datasets.
}\label{power_logis1}
}
\end{center}
\end{table}

\end{landscape}

\begin{example}
\normalfont
 \cite{self:1992} observed that the \cite{self:1988} method degrades when the sample sizes are highly unbalanced between two groups.
In this simulation, the set up is similar to that reported in Table 1 of  \cite{self:1992}. 
 We set
$\pi=\Pr(g=1)=0.05, 0.5, 0.75$, $\Pr(Z=2|g=1)=0.8$, and $\Pr(Z=2|g=0)=0.2$. Thus
$(\Pi_1,\Pi_2,\Pi_3,\Pi_4) =(0.8(1-\pi), 0.2(1-\pi), 0.2\pi, 0.8\pi)$. 
Note that  $\pi=0.5$ ($0.75$) corresponds to the $1:1$ ($3:1$) treatment  allocation ratio, which is commonly used in clinical trials.
The scenario $\pi=0.05$  may arise in case-control studies or in genetic studies when a small proportion of subjects carry the risk genotypes \citep{tang:2011}.
The true odds ratio is $\exp(\alpha_1)=2$ for  stratum and $\exp(\beta)=2$ for exposure. 
The overall response rate in the study population $$ \bar\mu=   \frac{\Pi_1}{1+\exp(-\alpha_0 )} +  \frac{ \Pi_2}{1+\exp(-\alpha_0 -\alpha_1  )} +     \frac{ \Pi_3 }{1+\exp(-\alpha_0  - \beta )}     +\frac{\Pi_4}{1+\exp(-\alpha_0 -\alpha_1- \beta )}$$
is set to $0.02$ and $0.15$. Because the sample size estimates vary greatly by methods, 
we evaluate the nominal power and empirical power based on $10^6$ simulations at both the sample sizes from the proposed  and \cite{self:1988} methods.  
We repeat the simulation  for unstratified score tests without adjustment for the stratum effect when there is no confounding effect ($\alpha_1=0$), where $\alpha_0$ is the solution to
$$ \bar\mu= \frac{\pi}{1+\exp(-\alpha_0  - \beta )}+ \frac{1-\pi}{1+\exp(-\alpha_0  )}.$$

Table \ref{power_logis2} displays the power and sample size results. The nominal power by formula \eqref{power} is very close to the simulated power, but formulae \eqref{size3} and \eqref{size1} may produce very poor power estimates which  can deviate from the simulated power by $16\%$ in some cases.
\end{example}

  \begin{table}[h]
\begin{center}{
\footnotesize
\begin{tabular}{c@{\extracolsep{5pt}}c@{\extracolsep{5pt}}c@{\extracolsep{5pt}}c@{\extracolsep{5pt}}c@{\extracolsep{5pt}}c@{\extracolsep{5pt}}c@{\extracolsep{5pt}}c@{\extracolsep{5pt}}c@{\extracolsep{5pt}}c@{\extracolsep{5pt}}c@{\extracolsep{5pt}}c@{\extracolsep{5pt}}c@{\extracolsep{5pt}}c@{\extracolsep{5pt}}c@{\extracolsep{5pt}}c@{\extracolsep{5pt}}c@{\extracolsep{5pt}}c@{\extracolsep{5pt}}c@{\extracolsep{5pt}}c@{\extracolsep{5pt}}c@{\extracolsep{5pt}}c@{\extracolsep{5pt}}c@{\extracolsep{5pt}}c@{\extracolsep{5pt}}rcccccccccccccccccccccccccc} \\\hline  
        &                   &    & target                & \multicolumn{3}{c}{estimated size}   & \multicolumn{4}{c}{ power ($\%$) at $N_{sw}$}  & \multicolumn{4}{c}{ power ($\%$) at $N_{new}$} \\ \cline{5-7}\cline{8-11}\cline{12-15}
$\pi$ & $\bar\mu$  & $\alpha_0$  & power($\%$)  & $N_{new}$ & $N_{sw}$ & $N_{s0}$   &  SIM  & $P_{new}$ & $P_{SM}$  & $P_{s0}$      &  SIM  & $P_{new}$ & $P_{SM}$  & $P_{s0}$              \\\hline 

\multicolumn{15}{c}{No confounding or stratum effect:  $p(y=1|z,g) = \frac{1}{1+\exp[-\alpha_0 -g\log(2)]}$}\\

                                                                                         $.05$&$.02$&$ -3.9398$&$  80$&$ 11661$&$ 17232$&$  9601$&$ 91.07$&$ 90.97$&$ 80.00$&$ 96.35$&$ 80.08$&$ 80.00$&$ 63.48$&$ 87.03$\\
                                                            &&&$  90$&$ 16537$&$ 23069$&$ 12853$&$ 96.27$&$ 96.23$&$ 90.00$&$ 99.14$&$ 90.06$&$ 90.00$&$ 78.36$&$ 95.70$\\
                                                            &$.15$&$ -1.7776$&$  80$&$  2035$&$  2626$&$  1805$&$ 88.37$&$ 88.04$&$ 80.00$&$ 92.22$&$ 80.02$&$ 80.00$&$ 69.37$&$ 84.51$\\
                                                            &&&$  90$&$  2826$&$  3516$&$  2416$&$ 94.99$&$ 94.71$&$ 90.01$&$ 97.45$&$ 90.27$&$ 90.00$&$ 82.80$&$ 93.90$\\
                                                            $.50$&$.02$&$ -4.2951$&$  80$&$  3587$&$  3581$&$  3589$&$ 80.63$&$ 79.95$&$ 80.01$&$ 79.92$&$ 80.69$&$ 80.01$&$ 80.07$&$ 79.99$\\
                                                            &&&$  90$&$  4800$&$  4794$&$  4804$&$ 90.55$&$ 89.97$&$ 90.00$&$ 89.94$&$ 90.59$&$ 90.00$&$ 90.04$&$ 89.98$\\
                                                            &$.15$&$ -2.1230$&$  80$&$   536$&$   531$&$   539$&$ 80.48$&$ 79.64$&$ 80.05$&$ 79.47$&$ 80.37$&$ 80.01$&$ 80.41$&$ 79.84$\\
                                                            &&&$  90$&$   717$&$   711$&$   721$&$ 90.18$&$ 89.78$&$ 90.04$&$ 89.62$&$ 90.42$&$ 90.03$&$ 90.28$&$ 89.86$\\
                                                            $.75$&$.02$&$ -4.4502$&$  80$&$  5879$&$  4651$&$  6451$&$ 69.37$&$ 68.91$&$ 80.00$&$ 66.24$&$ 80.73$&$ 80.00$&$ 88.30$&$ 76.26$\\
                                                            &&&$  90$&$  7636$&$  6226$&$  8636$&$ 83.20$&$ 82.47$&$ 90.00$&$ 78.59$&$ 90.46$&$ 90.00$&$ 94.84$&$ 86.17$\\
                                                            &$.15$&$ -2.2845$&$  80$&$   840$&$   697$&$   906$&$ 72.33$&$ 71.50$&$ 80.04$&$ 69.08$&$ 80.52$&$ 80.02$&$ 86.81$&$ 76.99$\\
                                                            &&&$  90$&$  1097$&$   933$&$  1213$&$ 84.85$&$ 84.33$&$ 90.03$&$ 81.17$&$ 90.24$&$ 90.00$&$ 94.02$&$ 86.95$\\

\\

\multicolumn{15}{c}{ Confounding: $p(y=1|z,g) = \frac{1}{1+\exp[-\alpha_0 -z\log(2)-g \log(2)]}$ }\\
                                                                                                         $.05$&$.02$&$ -4.1643$&$  80$&$  9752$&$ 13078$&$  8473$&$ 89.38$&$ 88.95$&$ 80.00$&$ 93.58$&$ 80.43$&$ 80.00$&$ 67.70$&$ 85.21$\\
                                                            &&&$  90$&$ 13621$&$ 17507$&$ 11343$&$ 95.48$&$ 95.19$&$ 90.00$&$ 98.06$&$ 90.41$&$ 90.00$&$ 81.57$&$ 94.43$\\
                                                            &$.15$&$ -1.9777$&$  80$&$  1943$&$  2269$&$  1811$&$ 85.58$&$ 85.33$&$ 80.01$&$ 88.02$&$ 80.30$&$ 80.00$&$ 73.66$&$ 82.69$\\
                                                            &&&$  90$&$  2659$&$  3037$&$  2425$&$ 93.54$&$ 93.19$&$ 90.00$&$ 95.24$&$ 90.39$&$ 90.01$&$ 85.84$&$ 92.43$\\
                                                            $.50$&$.02$&$ -4.7609$&$  80$&$  6096$&$  5596$&$  6317$&$ 76.89$&$ 76.40$&$ 80.00$&$ 75.08$&$ 80.46$&$ 80.00$&$ 83.25$&$ 78.59$\\
                                                            &&&$  90$&$  8068$&$  7492$&$  8456$&$ 88.09$&$ 87.68$&$ 90.00$&$ 86.24$&$ 90.37$&$ 90.00$&$ 91.98$&$ 88.62$\\
                                                            &$.15$&$ -2.5597$&$  80$&$   882$&$   853$&$   894$&$ 79.37$&$ 78.69$&$ 80.03$&$ 78.15$&$ 80.69$&$ 80.04$&$ 81.32$&$ 79.49$\\
                                                            &&&$  90$&$  1175$&$  1142$&$  1197$&$ 89.59$&$ 89.18$&$ 90.02$&$ 88.63$&$ 90.32$&$ 90.02$&$ 90.81$&$ 89.49$\\
                                                            $.75$&$.02$&$ -4.9868$&$  80$&$ 10398$&$  8795$&$ 11127$&$ 72.84$&$ 72.48$&$ 80.00$&$ 70.22$&$ 80.46$&$ 80.00$&$ 86.14$&$ 77.29$\\
                                                            &&&$  90$&$ 13618$&$ 11773$&$ 14896$&$ 85.45$&$ 85.01$&$ 90.00$&$ 82.17$&$ 90.42$&$ 90.00$&$ 93.65$&$ 87.27$\\
                                                            &$.15$&$ -2.8044$&$  80$&$  1419$&$  1271$&$  1486$&$ 76.01$&$ 75.27$&$ 80.03$&$ 73.62$&$ 80.46$&$ 80.00$&$ 84.17$&$ 78.18$\\
                                                            &&&$  90$&$  1872$&$  1701$&$  1989$&$ 87.46$&$ 86.92$&$ 90.01$&$ 85.04$&$ 90.35$&$ 90.00$&$ 92.52$&$ 88.20$\\

\hline
 \end{tabular} \caption{Power and sample size estimate  for the score test from  logistic regression with unequal group sample sizes \newline
[1] Simulated power (SIM) are evaluated at $N_{new}$ based on $1000,000$ simulated datasets.}\label{power_logis2}
}
\end{center}
\end{table}

\section{Discussion}
We propose a  modification of  the \cite{self:1988} method for sample size calculation for score tests from  GLMs, and extend it to the NI trials.
The modification takes into account of the fact that  the variance of the score statistic differs under $H_0$ and $H_1$.
The proposed method is also suitable for other regression models. For example,
the binary outcome is often analyzed by the logistic regression on basis of the odds ratio between two  groups.
Now suppose the parameter of interest is the relative risk instead of the odds ratio.  The  method is still suitable if the model is reparametrized in terms of
 the response rate in the control group and the relative risk parameter  \citep{tang:2019b}.

The proposed method shows a marked improvement over  the \cite{self:1988}  formula  in logistic and NB regressions when either the treatment effect is large or sample sizes are unbalanced in the two groups.
In these situations, the \cite{self:1988} method degrades because the variance of the score statistic can be quite different under $H_0$ and $H_1$.
As illustrated  in Section \ref{nbscore}, the  approaches of  \cite{zhu:2014} and  \cite{zhu:2017} for NB regression
tend to underestimate the  size 1) when there is a large variation in the patients' follow-up time, and/or 2) when there is a large treatment effect.

The sample size calculation for the score test requires the construction  and  analysis of  an exemplary dataset  if the model  is complex, and there is no analytic solution for the restricted MLE under the null hypothesis.
The main computation time lies in the analysis of the exemplary data, and this can usually be done within few minutes. It is much quicker than the simulation method, 
which requires the generation and analysis of at least thousands of datasets in order to get a quite precise power estimate at a given sample size.
 As evidenced by the results in Tables \ref{scorehigh} and \ref{power_sim1}, the sample size procedure shall be consistent with the test used for the analysis \citep{zhu:2014}.  
Otherwise, the study may be either underpowered or overpowered.
We recommend  using  the proposed   procedure (or a simplified version if it exists) 
to determine the sample size if one plans to analyze the trial using the score test.

A future research direction  is  to extend the  exemplary dataset  approach to the Wald test and score test \citep{liu:1997} from the  generalized estimating equations (GEE) in the analysis of repeated measurements. 
The sample size calculation is complicated even for the Wald test in GEEs because there are missing outcomes, and the working correlation structure may be different from the true correlation structure. 
Even if the analytic formula exists by using the independent or true  correlation structure, the calculation can still be complex  to account for missing data.
In the exemplary dataset approach,  one can get the noncentrality parameter for the Wald test directly through the analysis of the exemplary data. The generation of the exemplary data also provides an opportunity to verify whether the sample size assumption is correct for correlated outcomes.

\bibliographystyle{Chicago}

\bibliography{nbsizescore}

\end{document}